\documentstyle[preprint,aps,psfig]{revtex}

\tighten
\begin{document}
\draft
\preprint{TPI-MINN-99/19 $\;\;$
          UMN-TH-1752-99}

\newcommand{\nc}{\newcommand}
\nc{\al}{\alpha}
\nc{\ga}{\gamma}
\nc{\de}{\delta}
\nc{\ep}{\epsilon}
\nc{\ze}{\zeta}
\nc{\et}{\eta}
\renewcommand{\th}{\theta}
\nc{\Th}{\Theta}
\nc{\ka}{\kappa}
\nc{\la}{\lambda}
\nc{\rh}{\rho}
\nc{\si}{\sigma}
\nc{\ta}{\tau}
\nc{\up}{\upsilon}
\nc{\ph}{\phi}
\nc{\ch}{\chi}
\nc{\ps}{\psi}
\nc{\om}{\omega}
\nc{\Ga}{\Gamma}
\nc{\De}{\Delta}
\nc{\La}{\Lambda}
\nc{\Si}{\Sigma}
\nc{\Up}{\Upsilon}
\nc{\Ph}{\Phi}
\nc{\Ps}{\Psi}
\nc{\Om}{\Omega}
\nc{\ptl}{\partial}
\nc{\del}{\nabla}
\nc{\be}{\begin{eqnarray}}
\nc{\ee}{\end{eqnarray}}
\nc{\ov}{\overline}
\nc{\gsl}{\!\not}
\newcommand{\s}{\mbox{$\sigma$}}
\newcommand{\bi}[1]{\bibitem{#1}}
\newcommand{\fr}[2]{\frac{#1}{#2}}
\newcommand{\gm}{\mbox{$\gamma_{\mu}$}}
\newcommand{\gn}{\mbox{$\gamma_{\nu}$}}
\newcommand{\Le}{\mbox{$\fr{1+\gamma_5}{2}$}}
\newcommand{\R}{\mbox{$\fr{1-\gamma_5}{2}$}}
\newcommand{\GD}{\mbox{$\tilde{G}$}}
\newcommand{\gf}{\mbox{$\gamma_{5}$}}
\newcommand{\Ima}{\mbox{Im}}
\newcommand{\Rea}{\mbox{Re}}
\newcommand{\Tr}{\mbox{Tr}}

\title{The Theta Term in QCD Sum Rules and the Electric Dipole Moment of 
the Vector Meson}

\author{Maxim Pospelov\footnote{pospelov@mnhepw.hep.umn.edu} 
          and Adam Ritz\footnote{aritz@mnhepw.hep.umn.edu}\\}

\address{Theoretical Physics Institute, School of Physics and Astronomy \\
         University of Minnesota, 116 Church St., Minneapolis, MN 55455, USA}
\date{\today}

\maketitle

\begin{abstract}
We demonstrate that the QCD sum rule method can be successfully 
applied to the calculation of CP-odd electromagnetic observables 
induced by a vacuum $\theta$--angle. We implement the approach
in calculating the electric dipole moment of the rho meson
to $\sim 30$\% precision, and find that the result can also be explicitly 
related to the vacuum topological susceptibility.
\end{abstract}

\vfill\eject

\section{Introduction}
In this paper we demonstrate the feasibility of 
QCD sum rule calculations \cite{sr} for CP-odd 
electromagnetic observables induced by 
the QCD vacuum angle $\theta$. This parameter labels different
super-selection sectors for the QCD vacuum, and enters in front of the 
additional term in the QCD Lagrangian  
\be
{\cal L}= \theta\fr{g^2}{32\pi^2} G^a_{\mu\nu}\GD^a_{\mu\nu} 
\ee
which violates P and CP symmetries. As it is a total derivative,
$ G^a_{\mu\nu}\GD^a_{\mu\nu}$ can induce physical observables only through
non-perturbative effects. 

Experimental tests of CP symmetry suggest
that the $\th$--parameter is small, and among different CP-violating 
observables, the electric dipole moment (EDM) of the neutron is 
one of the most sensitive to the value of 
$\theta$ \cite{KL}. The calculation of
the neutron EDM induced by the theta term is a long standing problem. 
According to Ref.~\cite{CDVW}, an estimate of $d_n(\theta)$ can be obtained 
within chiral perturbation theory. The result,
\be
d_n=-e\theta\fr{m_u m_d}{f_\pi^2(m_u+m_d)}(\fr{0.9}{4\pi^2}
\ln(\Lambda/m_\pi) + c)
\label{eq:log}
\ee
is seemingly justified near the chiral limit where the 
logarithmic term becomes large and may dominate over 
other possible contributions parametrized in this formula by the constant $c$. 
This constant is not calculable within this formalism and in principle can 
be more important numerically than the logarithmic piece away from the
chiral limit. In fact it is also worth noting
that in the limit $m_u,~m_d \rightarrow 0$, the logarithm is
still finite, and stabilized, for example, 
by the electromagnetic mass difference between the proton and neutron. 
In any case, an inability to determine the size of corrections to the 
logarithm means that one is unable to estimate 
the uncertainty of this prediction.

If the logarithm is cut off at $\Lambda \sim m_\rho$  
and non-logarithmic terms are 
ignored, one can derive the following bound on the value of 
$\theta$ using current experimental results on the EDM of 
the neutron \cite{nEDM}: 
\be
\bar{\theta}<3\cdot10^{-10}.
\label{eq:dnt}
\ee
Confronted with a naive expectation $\theta\sim 1$, the experimental evidence
for a small if not zero value for $\theta$ constitutes a serious fine tuning 
problem, usually referred to as the strong CP problem.  The most popular 
solution for the strong CP problem is to allow the dynamical adjustment 
of $\theta$ to zero through the axion mechanism \cite{PQ}. 

There are two main motivations for improving the calculation of the EDM of the 
neutron induced by the theta term. The first refers to theories where the 
axion mechanism is absent and the $\theta$--parameter is zero at tree level 
as a result of exact P or CP symmetries \cite{P,CP}. 
At a certain mass scale these symmetries are 
spontaneously broken and a nonzero $\theta$ is induced through radiative 
corrections. At low energies, a radiatively induced theta term 
is the main source for the 
EDM of the neutron as other, higher dimensional, operators 
are negligibly small. As $\theta$ itself can be reliably calculated 
when the model is specified, the main uncertainty in predicting 
the EDM comes from the calculation of $d_n(\theta)$.

The second, and perhaps the dominant, incentive to refine the calculation
of $d_n(\theta)$ is due to efforts to
limit CP-violating phases in supersymmetric theories in general, and
in the Minimal Supersymmetric Standard Model (MSSM) in particular. 
Substantial CP-violating SUSY phases contribute significantly to $\theta$ and 
therefore these models apparently require the existence of the
axion mechanism. However, this does not mean that the $\theta$-parameter 
is identically zero. While removing $\theta\sim 1$, 
the axion vacuum will adjust itself to the minimum dictated by the 
presence of higher dimensional CP-violating operators which generate  
terms in the axionic potential linear in $\theta$. This induced
$\th$--parameter is then given by:
\begin{eqnarray}
\theta_{induced}=-\fr{K_1}{|K|}, \;\;\mbox{where}\;\;  \label{eq:k1}
K_1=i\left\{\int dx e^{iqx}\langle 0|T(\fr{\al_s}{8\pi}G\GD(x),
{\cal O}_{CP}(0)|0 \rangle\right\}_{q=0}\nonumber,
\end{eqnarray} 
where ${\cal O}_{CP}(0)$ can be any CP-violating operator with dim$>$4 
composed from quark and gluon fields, while 
\be
 K & = & i\left\{\int dx e^{iqx}\langle 0|T(\fr{\al_s}{8\pi}G\GD(x),
  \fr{\al_s}{8\pi}G\GD(0))|0 \rangle\right\}_{q=0} \label{K}
\ee
is the topological susceptibility correlator.
In the case of the MSSM, the most important 
operators of this kind are colour electric dipole moments of light quarks
$\bar{q}_igt^aG^a_{\mu\nu}\sigma_{\mu\nu}\gf q$, and three-gluon 
CP-violating operators. 
The topological susceptibility correlator $K$ was calculated in \cite{C,svz}
and the value of $\theta$, generated by color EDMs can be found in a 
similar way \cite{BUP}. Numerically, the contribution 
to the neutron EDM, induced by $\theta_{eff}$ is of the same order 
as direct contributions mediated by these 
operators and by the EDMs of quarks. Therefore, the complete calculation of 
$d_n$ as a function of the SUSY CP-violating phases must include  
a $d_n(\theta)$ contribution and a computation of this value, beyond the naive 
logarithmic estimate (\ref{eq:log}), is needed. 

Within the currently available techniques for the study of hadronic
physics, it seems that the only
chance to improve analytically on the estimate (\ref{eq:log}) is by use 
of the QCD sum rule method \cite{sr}. Given its success
in predicting various hadronic 
properties, including the electromagnetic form factors of baryons 
\cite{is,BY}, it appears highly suitable for the calculation of 
observables depending on $\theta$. In the sum rule approach,
physical properties of the 
hadronic resonances can be expressed through a combination of
perturbative and nonperturbative contributions, the latter parametrized in 
terms of vacuum quark-gluon condensates. In the case of CP-odd 
observables induced by $\theta$, the purely perturbative piece 
is absent and the result must be reducible to a set of the vacuum 
condensates taken in the electromagnetic and ``topological'' background.
The expansion to first order in $\theta$ will result in the appearance
of correlators which have a structure similar to $K$ and 
$K_1$ in Eq.~(\ref{eq:k1}). These correlators can then be calculated
via the use of current algebra, in a similar manner to those considered in 
\cite{C,svz}. In this approach the $\theta$--dependence 
will arise naturally, with the correct quark mass dependence, and 
the relation to the U(1) problem will be explicit. 
This relation is manifest in the 
vanishing of any $G\GD$-induced observable in the limit when the mass
of the U(1) ``Goldstone boson'' is set equal to the mass of pion. 

Obviously, the calculation of the EDM of the neutron induced by the theta term 
will be a substantial task, although it appears that the main problem
may be technical rather than conceptual -- the 
calculation of $d_n(\theta)$ needs to be at linear order in the 
quark mass as compared to the calculation of the anomalous magnetic moment
which may be performed in the limit $m_{u,d}=0$. 
There are, however, additional subtleties 
relating to the correct choice for the nucleon current in the presence of 
non-zero $\theta$. 

Keeping in mind the importance of a sum rule calculation for the EDM of 
the neutron, we would like to test the applicability of this method by 
calculating the EDM($\theta$) for a simpler system. The perfect candidate for 
this would be $\rho$ meson which couples to the isovector vector current
and whose properties have been predicted within QCD sum rules with 
impressive accuracy \cite{sr}. Thus in this paper we propose to study 
the feasibility of sum rule calculations for $\th$--dependent electromagnetic 
observables in this mesonic system. We begin, in Section 2, with a tree-level
analysis of the vector current correlator in a background with a nonzero
electromagnetic field, and a theta term. We obtain the Wilson OPE
coefficients for all $\th$--dependent contributions up to neglected operators
whose momentum dependence is $O(1/q^6)$. This result is also
briefly contrasted with the analogous expression for the tensor structure
leading to the magnetic dipole moment. In Section 3, we turn to
the phenomenological side of the Borel sum rule, and in Section 4 study
the various contributions to the sum rule in some detail, obtaining
a stable relation at the level of $O(1/q^4)$ terms, and a 
reasonably precise extraction of $d_{\rh}$. Section 5 contains
further discussion, including comments on a consistent procedure
for the definition of current operators away from the chiral limit
in a background with nonzero $\th$.

\section{Calculation of the Wilson OPE Coefficients}

Since it is a spin 1 particle, the $\rho$-meson can possess on-shell 
two CP-odd electromagnetic form factors, the electric dipole and
magnetic quadrupole moments. We shall concentrate here on the 
EDM of $\rho^{+(-)}$ as the CP-violating form factors of $\rho^{0}$,
induced by the theta term should vanish. This is a general consequence
of C-symmetry, which is respected by the theta term.

Before commencing the calculation, it is useful to have a
rough  estimate of the EDM of $\rho$ induced by theta. It is 
clear that the correct answer
should have the ``built-in'' feature of vanishing when 
$m_u$ or $m_d$ are sent to zero. Thus we should expect a result of the
form,
\be
d_\rho \sim \theta \fr{e}{m_\rho}\fr{m_um_d}{\Lambda (m_u+m_d)},
\label{estimate}
\ee
where $\Lambda$ is some scale at which the reduced quark mass is
effectively normalized, presumably between $f_\pi$ and $m_\rho$. 
We note in passing that one could also use the approach of
\cite{CDVW} to obtain the contribution to $d_\rho$ due to the chiral 
logarithm. 

In order to calculate the $\rh^+$ EDM within the sum rule approach, 
we need to consider the correlator
of currents with $\rh^+$ quantum numbers, in a background with nonzero
$\th$ and an electromagnetic field $F_{\mu\nu}$,
\be
 \Pi_{\mu\nu}(Q^2) & = & i\int d^4x e^{iq\cdot x}
    \langle 0|T\{j^+_{\mu}(x)j^-_{\nu}(0)\}|0\rangle_{\th,F},
\ee
where we denote $Q^2=-q^2$, with $q$ the current momentum.

To simplify the presentation, we shall consider the example
of $\rh^+$ in the $m_u=m_d$ limit, for which the current reduces
to $j_{\mu}^+=\ov{u}\ga_{\mu}d$. Since we always work to linear order in the
quark mass, it is straightforward to resurrect the full dependence when
required below, and we shall always write the full mass dependence explicitly,
with the implicit understanding that we set $m_u=m_d$. 
With this current structure, the correlator reduces
to the form
\be
 \Pi^+_{\mu\nu}(Q^2) & = & i\int d^4x e^{iq\cdot x}
    \langle 0|\ov{u}(x)\ga_{\mu}S^d(x,0)\ga_{\nu}u(0)
        +\ov{d}(0)\ga_{\nu}S^u(0,x)\ga_{\mu}d(x)|0\rangle_{\th,F}\nonumber\\
     & \equiv & \pi_{\mu\nu}^u+\pi_{\mu\nu}^d,
            \label{fullprop}
\ee
where we have contracted two of the quark lines leading to the
presence of the $d$ and $u$ quark propagators, $S^u(x,0)$ and $S^d(x,0)$, 
respectively.

We concentrate now on the contribution $\pi_{\mu\nu}^u$, and note that
at tree level the linear dependence on the background field $F_{\mu\nu}$ 
can arise either through a vacuum condensate, or from a vertex with 
the propagator. These contributions are depicted in Fig.~1, and correspond
to an expansion of the quark propagator to linear order in the
background field. If we assume a constant field $\ptl_{\rh}F_{\mu\nu}=0$,
the gauge potential may be written in covariant form
$A_{\mu}(x)=-\fr{1}{2}F_{\mu\nu}(0)x^{\nu}$, while similarly if we work in
a fixed point gauge \cite{smilga}, the gluon gauge potential may also be 
represented as $A_{\mu}^at^a(x)=-\fr{1}{2}G_{\mu\nu}^a(0)t^ax^{\nu}$.   
The expansion of the massless propagator, conveniently written
in momentum space, then takes the form \cite{svzrev}
\be
 S(q) & = & \int d^4x e^{iqx} S(x,0)\,=\, 
    \frac{1}{{\not\!q}}
         +\frac{q_{\al}}{(q^2)^2}e\tilde{F}_{\al\beta}\ga_{\beta}\ga_5
         +\frac{q_{\al}}{(q^2)^2}g\tilde{G}_{\al\beta}\ga_{\beta}\ga_5
            +\cdots, \label{propexp}
\ee 
where $G_{\mu\nu}=G_{\mu\nu}^at^a$, and we have introduced the dual field
strengths $\tilde{F}_{\al\beta}(=\frac{1}{2}\ep_{\al\beta\rh\si}F^{\rh\si})$,
and $\tilde{G}_{\al\beta}$. Since we are concerned only 
with the leading linear dependence on the quark mass, 
the mass structure of the propagator is very simple, and while not
shown explicitly here, this structure is easily resurrected when required.
The particular contributions we shall need will be given below.

While the expansion of the propagator (\ref{propexp}) 
apparently exhausts all possibilities
for obtaining a linear dependence on the background field, 
it is important to also consider
an expansion of the quark wavefunctions. The first order
correction in the covariant Taylor expansion will be sufficient here, and
is given by
\be
 u(x) & = & u(0)+x_{\al}D_{\al}u(0)+\cdots,  \label{wavefn}
\ee
where $D_{\al}=\ptl_{\al}-ie_u A_{\al}(x)$ is the covariant derivative in the
background field\footnote{In principle, next-to-leading order
contributions linearly proportional to the 
field strength may also arise from the second order term in the Taylor 
expansion. However, these contributions have a very small 
coefficient \cite{is}, and will be ignored, as they constitute a 
negligible correction to the terms we shall discuss below.}.

Thus, if we substitute the 
 expansions for the quark wavefunction and
propagator into $\pi_{\mu\nu}^u$, we find a sum of six terms,
\be
 \pi^u_{\mu\nu} & \equiv & i
    \left(\pi^u_1+\pi^u_2+\pi^u_3+\pi^u_{\ptl 1}+\pi^u_{\ptl 2}
                +\pi^u_{\ptl 3}\right), \label{sum}
\ee
which may be conveniently represented in momentum space as follows:
The first three terms, $\pi^u_{1,2,3}$, given by,
\begin{figure}
 \centerline{%
   \psfig{file=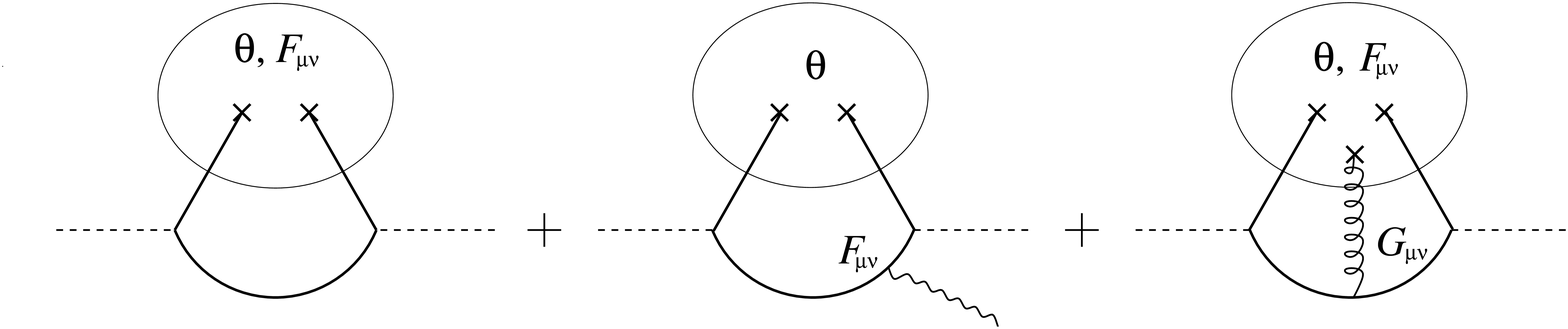,width=12cm,angle=0}%
         }
 \caption{Contributions to the correlator at leading order in $F_{\mu\nu}$.}
\end{figure}
\noindent
\be
 \pi^u_1 & = & 
    \langle 0|\ov{u}(0)\ga_{\mu}\frac{m_d}{(q^2)}\ga_{\nu}u(0)
                         |0\rangle_{\th,F} \label{p1}\\
 \pi^u_2 & = & 
      g\langle 0|\ov{u}(0)\ga_{\mu}\frac{m_d}{2(q^2)^2}(G\si)
                 \ga_{\nu}u(0)|0\rangle_{\th,F} \label{p2}\\
 \pi^u_3 & = & 
     e_d\langle 0|\ov{u}(0)\ga_{\mu}\frac{m_d}{2(q^2)^2}(F\si)
                 \ga_{\nu}u(0)|0\rangle_{\th} \label{p3}
\ee
represent the contributions 
at leading order in the quark wavefunction,
while $\pi^u_{\ptl 1,2,3}$ correspond to the first order corrections,
\be
 \pi^u_{\ptl 1} & = & -i\langle 0| \ov{u}(0)\stackrel{\leftarrow}{D_{\al}}
        \ga_{\mu}
 \left(\frac{g_{\al\beta}}{q^2}-2\frac{q_{\al}q_{\beta}}{(q^2)^2}\right)
     \ga_{\beta}\ga_{\nu}u(0)
                 |0\rangle_{\th,F} \label{pd1}\\
 \pi^u_{\ptl 2} & = & -ig\langle 0|\ov{u}(0)\stackrel{\leftarrow}{D_{\al}}
        \ga_{\mu}
 \left(\frac{g_{\al\beta}}{(q^2)^2}-4\frac{q_{\al}q_{\beta}}{(q^2)^3}\right)
           \tilde{G}_{\beta\ga}(0)\ga_{\ga}\ga_5
       \ga_{\nu}u(0)|0\rangle_{\th,F} \label{pd2}\\
 \pi^u_{\ptl 3} & = & -ie_d\langle 0|\ov{u}(0)\stackrel{\leftarrow}{D_{\al}}
        \ga_{\mu}
 \left(\frac{g_{\al\beta}}{(q^2)^2}-4\frac{q_{\al}q_{\beta}}{(q^2)^3}\right)
           \tilde{F}_{\beta\ga}(0)\ga_{\ga}\ga_5
       \ga_{\nu}u(0)|0\rangle_{F}. \label{pd3}
\ee
Note that in evaluation of $\pi^u_2$ and $\pi^u_{\ptl 2}$
we may turn off the background electromagnetic field.

We shall now consider each of these contributions in turn, although
as a first step its helpful to re-express the quark wavefunction
corrections $\pi^u_{\ptl 1,2,3}$ in terms of the leading order
condensates via use of the equations of motion.
It proves convenient to first analyze the derivative term $\pi^u_{\ptl 1}$. 
Writing this result in spinor notation,
so as to factorise the $\ga$--matrix structure, the matrix element we need
to consider has the form\footnote{In principle one could also explicitly 
include colour indices. However, the trace over these indices will always
be trivial in the examples to be considered here, so this dependence
will be suppressed.} $\langle 0| \ov{u}_a \stackrel{\leftarrow}{D_{\al}}
u_b |0\rangle$. Since we are only concerned with matrix elements which give
a nonzero contribution when evaluated in the $\th$--vacuum, its helpful
to choose an appropriate basis of condensates in which to expand.
For this example, there are two natural vector and axial-vector structures
to consider. We write
\be
 \langle 0| \ov{u}_a D_{\al}
         u_b |0\rangle & = & C_1(\ga_{\beta})_{ba}\langle 0| \ov{u}
          \ep_{\al\beta\mu\la}D_{\mu}\ga_{\la}\ga_5 u| 0\rangle \nonumber\\
       & & \;\;\;\;\; +C_2(\ga_{\beta}\ga_5)_{ba}\langle 0| \ov{u}
          D_{[\al}\ga_{\beta]}\ga_5 u| 0\rangle, \label{decomp}
\ee 
where $C_1$ and $C_2$ are two constants to be determined and $[\al,\beta]$
is used to denote anti-symmetrisation of indices (symmetrisation of indices
will later be denoted by $\{\al,\beta\}$).

Making use of the following identities,
\be
 \{\not\!\!D,\si_{\al\beta}\} & = & -2\ep^{\al\beta\mu\la}\ga^{\la}
                   \ga_5D_{\mu}\\
 \left[\not\!\!D,\si_{\al\beta}\right]\ga_5 & = & 2iD_{[\al}\ga_{\beta]}
                    \ga_5,
\ee
integrating by parts, and using the Dirac equation $\not\!\! D u =-im_u u$,
we can reduce the natural Lorentz decomposition (\ref{decomp}) to a 
more recognizable form,
\be
 \langle 0| \ov{u}_a D_{\al}
         u_b |0\rangle & = & im_u C_1(\ga_{\beta})_{ba}\langle 0| \ov{u}
             \si_{\al\beta}u|0\rangle \nonumber \\
      & & \;\;\;\; -m_uC_2(\ga_{\beta}\ga_5)_{ba}\langle 0| \ov{u}
              \si_{\al\beta}\ga_5 u|0\rangle.
\ee
Two equations for the constants $C_2$ and $C_2$ may be obtained by,
in one case, contracting with
$(\si_{\mu\nu}\ga_{\al})_{ab}$, and in another, via 
multiplication by $\ga_{\de}$ and
then anti-symmetrising in $\al,\de$. One obtains $C_1=0$ and $C_2=-1/8$,
and thus we are left with only one structure in the decomposition.

On Fourier transforming to momentum space, and performing the straightforward
$\ga$--matrix algebra we find
\be
  \pi^u_{\ptl 1} & = &
         i\frac{m_d}{q^2}\langle 0| \ov{u}\si_{\mu\nu}u|0\rangle_{\th,F}
     -m_d\frac{q_{\al}q_{\ga}}{(q^2)^2}\ep_{\mu\nu\ga\beta}
          \langle 0|\ov{u}\si_{\al\beta}\ga_5 u|0\rangle_{\th,F}.
\ee
We can now compare the first term here with $\pi^u_1$ given in
(\ref{p1}). Since the $\th$--dependent CP-odd contribution corresponds to
considering only the term antisymmetric in $\mu$ and $\nu$, the
relation $\ga_{\mu}\ga_{\nu}=-i\si_{\mu\nu}+$sym. implies that $\pi^u_1$
precisely cancels the first term above. Thus we have,
\be
  \pi^u_1+\pi^u_{\ptl 1} & = &
     -m_d\frac{q_{\al}q_{\ga}}{(q^2)^2}\ep_{\mu\nu\ga\beta}
          \langle 0|\ov{u}\si_{\al\beta}\ga_5 u|0\rangle_{\th,F}.
                  \label{p12}
\ee

Turning next to $\pi^u_{\ptl 2}$ (\ref{pd2}), a little $\ga$--matrix
algebra shows that the matrix element may be rewritten in the
form
\be
 \pi^u_{\ptl 2} & = & \left(\frac{g_{\al\rh}}{(q^2)^2}
    -4\frac{q_{\al}q_{\rh}}{(q^2)^3}\right)\ep_{\mu\si\nu\la}
   \tilde{F}_{\rh\si}
    \langle 0|\ov{u}\stackrel{\leftarrow}{D_{\al}}\ga_{\la}u|0\rangle_{\th}.
\ee
It can be shown that the condensate in this expression is in fact 
proportional only to $g_{\alpha\lambda}m_u\langle 0|\ov{u}u|0\rangle_{\th}$
and therefore does not contain any CP-violating piece.

The final wavefunction correction to consider is $\pi^u_{\ptl 3}$
(\ref{pd3}) which may be handled in a similar manner to $\pi^u_{\ptl 1}$,
via extracting the appropriate projections onto vacuum condensates, or
alternatively by direct calculation. We shall follow the former approach
here, and write down
\be
 \langle 0|\ov{u}_p (\tilde{G}_{\beta\ga}^a)
  D_{\al}u_q|0\rangle & = & C_1(t^a)(\ga_{\al})_{pq}
  \langle 0|\ov{u} (\tilde{G}_{\beta\ga})\not\!\!D u|0\rangle.
\ee
Note that another apparently valid Lorentz structure of the form
$\langle 0|\ov{u} (\tilde{G}_{\beta\ga}^a)\ga_5\not\!\!D u|0\rangle$
vanishes on the equations of motion.
Contracting with $(\ga_{\al})_{qp}(t^a)_{ji}$, and recalling 
that $t^at^a=4/3$, one finds that $C_1=3/64$. The 
resulting expression for $\pi^u_{\ptl 3}$ is,
\be
 \pi^u_{\ptl 3} & = & -i\frac{m_u g}{4}
 \left(\frac{g_{\al\beta}}{(q^2)^2}-4\frac{q_{\al}q_{\beta}}{(q^2)^3}\right)
 \ep_{\mu\nu\al\ga}\langle 0|\ov{u}\tilde{G}_{\beta\ga}u|0\rangle.
\ee

The next problem to address is that of extracting the leading $\th$--dependence
of these matrix elements, and we follow standard practice 
(see e.g. \cite{svz}) in making use of the anomalous Ward identity.
To illustrate the procedure, consider the condensate 
$m_u\langle 0|\ov{u}\Ga u|0\rangle$, with a generic Lorentz structure
denoted by $\Ga$. In the $\th$--vacuum, to leading order, we have
\be
 m_u\langle 0|\ov{u}\Ga u|0\rangle_{\th}
      & = & m_u\int d^4 y\langle 0|T\{(\ov{u}\Ga u(0),i\th\frac{\al_s}
        {8\pi}G^a_{\mu\nu}\tilde{G}^{a\mu\nu}(y)\}|0\rangle.
               \label{theta}
\ee
We now make use of the anomalous Ward identity for axial currents
\cite{svz} restricted to 2 flavours. A useful calculational
simplification follows if we take as the anomaly relation a linear 
combination of the singlet equations for the $u$ and $d$ quarks.
In particular, we use
\be
 \ptl_{\mu}j_{\mu 5} & = & 2m_*(\ov{u}i\ga_5u+\ov{d}i\ga_5d)
      +\frac{\al_s}{4\pi}
        G_{\mu\nu}^a\tilde{G}^{a\mu\nu}, \label{anom}
\ee
where 
\be
 j_{\mu 5} & = & \frac{m_*}{m_u}\ov{u}\ga_{\mu}\ga_5u
                    +\frac{m_*}{m_d}\ov{d}\ga_{\mu}\ga_5d
\ee
is the anomalous current, and we have introduced the reduced mass,
\be
 m_* & \equiv & \frac{m_u m_d}{m_u+m_d}.
\ee

Substituting the anomaly relation (\ref{anom}) for $G\tilde{G}$ into
the correlator (\ref{theta}), we recall that 
the only contribution
from $\ptl_{\mu}j_{\mu 5}$ is a contact term $\propto \de(y_0)$ due to the 
presence of the $T$--product. Through the use of the equal time   
commutator, we find that this leads to a local contribution
(independent of $y$), and consequently we have
\be
  m\langle 0|\ov{u}\Ga u|0\rangle_{\th} & = & i\th m_u \frac{m_*}{m_u}
      \langle 0| \ov{u}\Ga\ga_5 u|0\rangle \nonumber\\
 & & \;\;\;\;\;\;\;\;\;\;
     +i\th\int d^4y\langle 0|T\{m_*(\ov{u}\ga_5u(y)+\ov{d}\ga_5d(y)),
               m_u\ov{u}\si_{\mu\nu}u(0)\}|0\rangle.
\ee
The nonlocal contribution to this correlator, the second term above,
is $O(m^2)$ in light quark masses. Nonetheless, this term would cancel the
local contribution were there an intermediate state with mass squared
of $O(m)$ -- for example the Goldstone boson in the singlet channel. 
The crucial point, as stressed in \cite{svz}, is that
due to the $U(1)$ problem the lightest intermediate state $\et$ has 
$m_{\et}\gg m_{\pi}$ and thus the second term can be neglected at leading
order in $m$. Thus for each of the contributions above, the leading
dependence on $\th$ is determined via the following relation,
\be
 m\langle 0|\ov{u}\Ga u|0\rangle_{\th} & = & i\th m_*
      \langle 0| \ov{u}\Ga\ga_5 u|0\rangle. \label{thetrel}
\ee

We could of course have obtained this result in a simple manner by using
the anomaly to rotate away the $G\tilde{G}$ term in the action.
This induces a complex quark mass $m\rightarrow m+i\th m_{*}\ga_5$, and leads
directly to the leading $\th$--dependence (\ref{thetrel}) above.
However, despite being somewhat more involved, 
the procedure we have followed is advantageous in that it makes quite
explicit the role of the anomaly and, in particular, the conditions under which
the higher order non-local terms may be neglected. We shall return to the
issue of chiral rotations at the level of the action in Section~5.

The final effect to consider is that of the background field, $F_{\mu\nu}$.
For the term $\pi^u_{3}$ (\ref{p3}), the leading $F$-dependence
is already explicit, and may be extracted via introduction of spinor
indices. For the other terms, we follow Ioffe and Smilga \cite{is}
and introduce ``condensate susceptibilities'', $\ch$, $\ka$, and $\xi$,
defined as follows \cite{is}:
\be
 \langle 0| \ov{q}\si_{\mu\nu}q|0\rangle_F & = & e_q\ch F_{\mu\nu}
           \langle 0| \ov{q}q|0\rangle \nonumber\\
 g\langle 0| \ov{q}(G_{\mu\nu}^at^a)q|0\rangle_F & = & e_q\ka F_{\mu\nu}
           \langle 0| \ov{q}q|0\rangle \label{Frel}\\
 g\ep_{\mu\nu\la\si}\langle 0| \ov{q}\ga_5(G_{\la\si}^at^a)q|0\rangle_F 
   & = & ie_q\xi F_{\mu\nu}\langle 0| \ov{q}q|0\rangle \nonumber,
\ee
where $q=u$ or $d$.

Using the relations (\ref{thetrel},\ref{Frel}), and performing the 
Fourier transformation to
momentum space, we can now gather all the results from
$\pi^u_1$--$\pi^u_{\ptl 3}$ (\ref{sum}) and combine them with the
analogous results for $\pi^d_{\mu\nu}$ in (\ref{fullprop}), to obtain
\be
 \Pi^+_{\mu\nu} & = & m_*\th(e_u-e_d)\langle 0|\ov{q}q|0\rangle
     \left[\frac{\tilde{F}_{\mu\nu}}{q^2}\left(-\ch-\frac{1}{q^2}
      \left(1+\ka-\frac{1}{4}\xi\right)\right)\right. \nonumber\\
   & & \;\;\;\;\;\;\;\;\;
       -\left.\left(\ch-\frac{\xi}{q^2}\right)\frac{q_{\al}q_{[\mu}
      \tilde{F}_{\nu]\al}}{(q^2)^2}\right].  \label{opefull}
\ee
This expression exhibits the two tensor structures one would have
expected to appear on general grounds. However, only the first term
contributes to the EDM, as one may check by choosing a rest frame for the 
current momentum, since the second tensor structure vanishes
on shell.

Therefore, if we retain only the contribution which survives on-shell,
we have as our final expression for the theoretical side of sum rule,
\be
 \Pi^+_{\mu\nu} & = & m_*\th(e_u-e_d)\langle 0|\ov{q}q|0\rangle
    \left[\frac{\tilde{F}_{\mu\nu}}{q^2}
     \left(-\ch-\frac{1}{q^2}\left(1+\ka-\frac{\xi}{4}\right)\right)\right].
           \label{opefinal}
\ee

As a short digression, it is
instructive to contrast this result with the analogous expression
one would obtain for the structure $F_{\mu\nu}$ which leads to an
extraction of the magnetic dipole moment $\mu_{\rh}$. The crucial
difference is that in this case a nonzero contribution survives in the
chiral limit. Specifically, a perturbative 1-loop diagram in the background
field leads to a contribution of the form,
\be
 \Pi^+_{\mu\nu} & = & -\frac{1}{8\pi^2}
               (e_u-e_d)F_{\mu\nu}\ln \frac{\La^2}{-q^2}+\cdots.
             \label{Pi_mu}
\ee
Subleading power corrections have been ignored here. However,
it turns out that such contributions generically have a form similar
to those in (\ref{opefinal}) with coefficients $m_*\th\rightarrow m_q$,
and therefore vanish in the chiral limit $m_q\rightarrow 0$. 
One may also check that the first subleading terms of $O(m_q^0)$ actually
vanish identically, and therefore the perturbative piece serves
as the dominant contribution to $\mu_{\rh}$. An interesting corollary
is that, while certain power corrections are closely related as above,
there would appear to be no simple proportionality relation
between the electric and magnetic dipole moments.

\section{Mesonic Spectral Function and Construction of the Sum Rule}

In order to extract a numerical value for the $\rh^+$ EDM from the
OPE, we assume as usual that $\Pi^+$ satisfies a dispersion relation
(ignoring subtractions) of the form, 
\be
 \Pi(q^2) & = & \frac{1}{\pi}\int_0^{\infty}d\si \frac{Im\Pi(\si)}{(\si-q^2)},
\ee
which we then saturate with physical mesonic states ($\rh^+$, and excited
states with the same quantum numbers which we denote collectively as $\rh'$). 
To suppress the contribution
of excited states, we apply a Borel transform to $\Pi^+$, which we define,
following \cite{svzrev,rry}, as
\be
 {\cal B}\Pi^+ & \equiv & \mbox{lim}_{s,n\rightarrow\infty,s/n=M^2}
                    \frac{s^n}{(n-1)!}\left(-\frac{d}{ds}\right)^n\Pi^+(s)
           = \frac{1}{\pi M^2}\int_0^{\infty}d\si e^{-\si/M^2}Im\Pi^+(\si),
\ee 
where $s=-q^2$.

The phenomenological side of the sum rule may be parametrised by considering
the form-factor Lagrangian which encodes the effective CP violating
vertices (see Fig.~2). This has the form 
${\cal L}=\sum_n f_nS(q){\cal O}_nS(q)$, where
$f_n$ is the form factor, $S(q)$ is the on-shell propagator for $\rh^+$
or one of its excited states, and ${\cal O}_n$ is the operator 
corresponding to the induced vertex.

As mentioned above, there are a priori two such 
operators which need to be considered at lowest
order, $\tilde{F_{\mu\nu}}q^2$ and $q_{\al}q_{[\mu}\tilde{F}_{\nu]\al}$. 
As noted above, the second structure vanishes on-shell and thus does
not enter the form-factor Lagrangian. Another on-shell T-odd form factor,
the magnetic quadrupole moment, would appear only at the next order in 
momentum transfer, i.e. in front of the structure proportional to 
$\partial_\lambda F_{\mu\nu}$ \cite{KP}. Since we work only to linear 
order in photon momentum, the magnetic quadrupole moment 
cannot be recovered from the OPE form (\ref{opefinal}), and so we omit it 
on the phenomenological side as well.
\begin{figure}
 \centerline{%
   \psfig{file=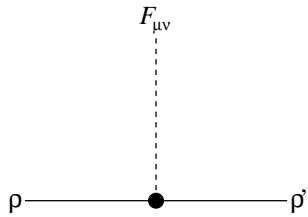,width=4cm,angle=0}%
         }
 \hspace{0.1in}
 \caption{Mesonic contributions to the current correlator in an external
 electromagnetic field. Possible excited states with the $\rh^+$ quantum
numbers are denoted generically by $\rh'$.}
\end{figure}
\noindent
Consequently, for comparison with the OPE,
we have on the phenomenological side in momentum space,
\be
 \Pi^{+(phen)}_{\mu\nu} & = & 2f(q^2)\tilde{F}_{\mu\nu} +\cdots, 
          \label{phenfull}
\ee
where, since we work outside the dispersion relation we may add polynomials
in $q^2$ to ensure transversality in the chiral limit, 
and optimum behaviour for large $q^2$, without affecting the physical 
spectral function $Im\Pi$. We then find that the 
function $f(q^2)$ takes the form,
\be
 f(q^2) & = & \frac{f_1\la^2}
     {(q^2-m_{\rh}^2)^2}+\sum_{n}\frac{f_n}{(q^2-m_{\rh}^2)
    (q^2-m_n^2)}+\sum_{n,m}\frac{f_{nm}}{(q^2-m_n^2)(q^2-m_m^2)}.
\ee
In this expression, $\la$ is the dimension 2 coupling defined in terms 
of the transition amplitude
for the vector current to go into the $\rh^+$ state, 
$\langle 0|j_{\mu}|\rh^+\rangle=\la V_{\mu}$, where $V_{\mu}$ is an
appropriate vector, 
while $f_1$ is associated with the $\rh^+$ EDM, and $f_n$ and $f_{nm}$
correspond respectively to transitions between 
$\rh^+$ and excited states, and between
the excited states themselves. 

After performing a Borel transform on $f$, we obtain\footnote{An alternative
derivation of the Borel transform in this context, using double dispersion
relations in order to parametrise $Im\Pi$, was presented in \cite{is}.}
\be
 f(M^2) & = & \frac{\la^2 f_1}{M^4}e^{-m_{\rh}^2/M^2}
     +\frac{1}{M^2}\sum_n\frac{f_n}{m_n^2-m_{\rh}^2}
                e^{-m_{\rh}^2/M^2}\nonumber\\
   & &  \;\;\;\;\;\;\;\;\;\;+\sum_{n,m}\frac{f_{nm}}{(m_n^2-m_{\rh}^2)
        (m_m^2-m_{\rh}^2)}
       e^{-(m_n^2+m_m^2)/M^2}. \label{f}
\ee
Since the gap from $m_{\rh}\sim 0.77$GeV, to the first excited state
$m_{\rh'}\sim 1.7$GeV is large, we shall, as in \cite{kw}, ignore the 
continuum contribution as it is exponentially suppressed. Thus we may
write,
\be
 f(M^2) & = & \left(\frac{\la^2 f_1}{M^4}+\frac{A}{M^2}\right)
                       e^{-m_{\rh}^2/M^2}, \label{fA}
\ee
where $A$ is an effective constant of dimension 2.

We are now in a position to write down the sum rule for the coefficient
of $\tilde{F}_{\mu\nu}$. From the Borel transform
of (\ref{opefinal}), and also (\ref{phenfull},\ref{fA}), we have
\be
 \la^2 f_1+AM^2 & = & \frac{1}{2}m_*\th(e_u-e_d)M^4e^{m_{\rh}^2/M^2}
    \langle 0|\ov{q}q|0\rangle
     \left(\frac{\ch}{M^2}-\frac{1}{M^4}\left(1+\ka-\frac{\xi}{4}\right)
              +\cdots\right). \label{sumrule}
\ee  
This is our final result for the CP-odd sum rule, and will be investigated
numerically in the next section.

\section{Numerical Analysis}

The coupling $\la$ present in (\ref{sumrule}) may be obtained
from the well known mass sum rule in the CP even sector. In this case, there
is no need to consider a background electromagnetic field, and the sum
rule takes the form (see e.g. \cite{rry}),
\be
 \frac{\la^2}{m_{\rh}^2M^2}e^{-m_{\rh}^2/M^2} & = & -\frac{1}{4\pi^2}\left(1+
     \frac{\al_s}{\pi}\right)e^{-s_0/M^2} \nonumber\\
    && \;\;\;\;\;\;
    +\frac{1}{4\pi^2}\left(1+\frac{\al_s}{\pi}+\frac{\pi^2}{3M^4}\left(
     \langle 0|\frac{\al_s}{\pi}G^2 |0\rangle+24\langle 0|m_q \ov{q}q|0\rangle
    \right)\right). \label{evenSR}
\ee
Since there is no background field, the leading term is a single pole
contribution, and we include a continuum term shifted to the
right-hand side starting from the $\rh'$ threshold at $s_0\sim 1.7$GeV.
Note also that $\la$, as defined above, is related to the dimensionless
coupling $g_{\rh}$ associated with the width of the resonance, via
$\la=m_{\rh}^2/g_{\rh}$, so that $g_{\rh}$ is dimensionless.

The physical EDM parameter $d_{\rh}$ may be obtained by
normalising the form factor $f_1$, introduced above, by the
$\rh^+$ mass. Furthermore, it will be convenient in what follows
to define an additional parameter $\tilde{d}$ via the relation,
\be 
 d_{\rh} & = & \frac{f_1}{m_{\rh}}\; \equiv \; \tilde{d} 
             \frac{m_*}{m_{\rh}}\th(e_u-e_d).
\ee

We shall now study the sum rule (\ref{sumrule}), making use of
(\ref{evenSR}) to remove the $\la$--dependence. Its helpful to
consider the various contributions to (\ref{sumrule}) in turn.
At the most naive level, we can ignore the $O(1/M^4)$ corrections
in (\ref{sumrule}), and also the continuum in (\ref{evenSR}). Taking
ratios one finds,
\be
 \tilde{d}_1 & \sim & \frac{2\pi^2}{m_{\rh}^2} 
            \frac{\ch\langle 0|\ov{q}q|0\rangle}{(1+\al_s/\pi
            +\pi^2\langle {\cal O}_4\rangle/M^4)}, \label{naive}
\ee
where we have defined $\langle {\cal O}_4\rangle\equiv\langle 0|
\frac{\al_s}{\pi}G^2 |0\rangle+24\langle 0|m_q \ov{q}q|0\rangle$ for 
convenience. For numerical calculation we make use of the following
parameter values: For the quark condensate, we have
\be
 \langle 0|\ov{q}q|0\rangle & = - (0.225\mbox{ GeV})^3,
\ee
while for the condensate susceptibilities, we have the values calculated
in \cite{chival} and \cite{kw},
\be
 \ch & = & - 5.7 \pm 0.6 \mbox{ GeV}^{-2} \mbox{\cite{chival}} \\
 \ka & = & - 0.34 \pm 0.1  \mbox{\cite{kw}} \\
 \xi & = & - 0.74 \pm 0.2  \mbox{\cite{kw}}
\ee
Note that $\ch$, which enters at $O(1/M^2)$, since it is dimensionful,
is numerically significantly
larger than $\ka$ and $\xi$. With these parameters, the result 
for $\tilde{d}_1$ is shown in Fig.~3, where
we have also used the 1-loop running coupling $\al_s(M)$ with two
flavours, normalised to $0.34$ at $M_{\ta}$. Note that
the stability at large $M^2$ is an artefact of the cancelation of the
leading $M$ dependence in (\ref{naive}). One should expect this relation
to have reasonable accuracy only in the range $M^2\sim m_{\rh}^2$.

To observe the effect of the $O(1/M^4)$ corrections, we now need to
address the issue of the unknown constant $A$ in (\ref{sumrule}).
Relative to $f_1$, there is a suppression factor of $M^2/(s_0-m_{\rh}^2)$
associated with $A$, which near $M^2=m_{\rho}^2$ is $\sim 0.25$. 
Although this is to be summed over all the excited states, we shall
use this as justification to treat $A$ perturbatively, and solve 
for it in terms of $f_1$ using (\ref{sumrule}), but ignoring
$O(1/M^4)$ terms. Its convenient to do this via first
pre-multiplying (\ref{sumrule}) by $M^2$ and then differentiating
by $1/M^2$. One obtains the relation
\be
 A & \sim & f_1\la^2\left(\frac{1}{m_{\rh}^2}-\frac{1}{M^2}\right)+\cdots,
\ee 

\begin{figure}
 \centerline{%
   \psfig{file=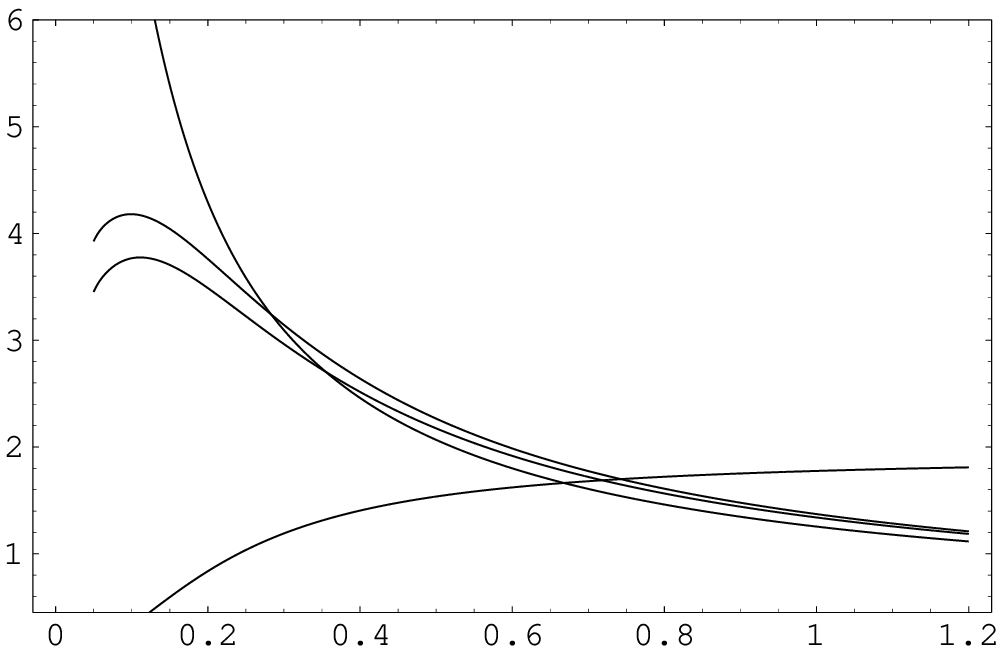,width=10cm,angle=0}%
         }
 \vspace*{-3.5cm}\hspace*{0.5cm} $\tilde{d}$(GeV$^{-1}$)

 \vspace*{-3cm}\hspace*{4.6cm} $\tilde{d}_2$

 \vspace*{2.7cm}\hspace*{11cm} $\tilde{d}_1$

 \vspace*{-3.0cm}\hspace*{3.5cm} $\tilde{d}_3$

 \vspace*{1.06cm}\hspace*{3.7cm} $\tilde{d}_4$

 \vspace*{2.6cm}\hspace*{12.7cm} $M^2$(GeV$^2$)

 \vspace*{0.5cm}

 \caption{The $\rh^+$ EDM parameter $\tilde{d}$
as a function of $M^2$ according to various 
components of the sum rules (\ref{naive}) and (\ref{dtilde}).}
\end{figure}

\noindent
which vanishes when $M^2=m_{\rh}^2$. Substituting this back into
(\ref{sumrule}), we can isolate $\tilde{d}$ by taking the ratio
with the quantity obtained by pre-multiplying (\ref{evenSR})
by $e^{s_0/M^2}$, and then differentiating by $1/M^2$,
\be
 \tilde{d} & = & 2\pi^2\frac{\langle 0|\ov{q}q|0\rangle(s_0+M^2-m_{\rh}^2)}
   {(s_0(1+\al_s/\pi)+\pi^2(s_0+2M^2)\langle {\cal O}_4\rangle/M^4)}
 \left(\frac{\ch}{M^2}-\frac{1}{M^4}\left(1+\ka-\frac{\xi}{4}\right)
              +\cdots\right). \label{dtilde}
\ee

To study the various contributions to (\ref{dtilde}), we first
set all the $O(1/M^4)$ corrections zero, and the result, denoted
$\tilde{d}_2$, is shown in Fig.~3. We see the expected $1/M^2$ behaviour
so that there is no stability region, although the relation is, as one
would expect,
very close to $\tilde{d}_1$ (\ref{naive}) near $M^2=m_{\rh}^2$. 

The leading correction at $O(1/M^4)$ may be isolated by setting
$\ka=\xi=0$ in (\ref{dtilde}). The result, $\tilde{d}_3$, is shown
in Fig.~3. The presence of the $1/M^4$ term induces a transition
region in the $M^2$ dependence. Note that the numerical similarity with
$\tilde{d}_2$ over the relevant range of the Borel parameter $M^2$ is in 
part due to the compensating effect of the $O(1/M^4)$ correction
in the denominator of (\ref{dtilde}).

Finally, we can obtain an estimate of the corrections associated with
$\ka$ and $\xi$, by plotting the full expression in (\ref{dtilde}),
which is displayed in Fig.~3 as $\tilde{d}_4$. We see that
including these corrections has little effect on the behaviour or stability
of the sum rule. This is encouraging, as the precise numerical values
for $\ka$ and $\xi$ are uncertain to a larger degree than that of $\ch$. 
Extracting a numerical estimate for $\tilde{d}$, 
and an approximate error, from the
stability region $M^2\sim 0.3-0.8$GeV$^2$ in Fig.~3, we find the result
\be
 d_{\rh} & = & (2.6 \pm 0.8)e\th\fr{m_*}{({\rm 1 GeV})^2},
\label{final}
\ee
for the EDM of $\rh^+$, where $e=e_u-e_d$ is the positron charge.
As is clear from Fig.~3, the dominant contribution
arises from the term proportional to the susceptibility $\ch$, and thus
the result is essentially linearly dependent on the value (and error) 
for this parameter. 

It is interesting to note that comparison with the naive 
estimate (\ref{estimate}) implies a value for the effective scale $\Lambda$
of order 1 GeV. This is very close to a similarly defined scale which
would effectively appear in the chiral logarithm 
estimates for $d_n$, Eq. (\ref{eq:log}). In this sense, we can conclude 
that our result is in the expected range. One final point to note
is that if we return to Eq.~(\ref{dtilde}), which is written in terms of the 
condensates, and re-express the final answer in slightly different units,
we can directly relate the EDM to the vacuum topological susceptibility 
correlator $K$ (\ref{K}) as calculated in \cite{C,svz}: 
\be
 d_{\rh} & = & 2.2\times 10^{-3}e\th\fr{m_\pi^2 f_\pi^2}{({\rm 100 MeV})^5}
\fr{m_u m_d}{(m_u+m_d)^2}=2.2\times 10^{-3}e\th\fr{K}{({\rm 100 MeV})^5}.
\ee
Note that we have used a normalisation (100MeV) which is adapted 
to the small size of $m_*$, and it is this which accounts for the 
small overall coefficient. This factor is essentially hidden 
in the result presented in (\ref{final}).

\section{Discussion}

Throughout the calculation we have intentionally kept $m_u=m_d$, knowing that 
the correct mass dependence is $m_*$. It is easy to see, however, that 
if $m_u\neq m_d$ the calculation does not automatically 
restore $m_*$ since, for example, the up-quark bilinear combination in 
Eq.~(\ref{p3}) comes with a
coefficient proportional to $m_d$. In the final result it would induce 
a contribution which would not vanish in the limit $m_u\rightarrow 0$. This 
means that if $m_u\neq m_d$ there should be additional contributions
which would combine with the rest to form an overall $m_*$-dependent result.

At the same time, we recall that one can use a chiral transformation 
in the QCD Lagrangian and rotate
the $\theta$-parameter to stand in front of the quark singlet combination
$m_*(\bar u i\gf u + \bar d i\gf d)$. It is clear that in this situation
the $\theta$-dependence for any physical observable will arise 
together with the 
correct mass dependence and will disappear at $m_u=0$. The answer to this 
``puzzle'' lies in the chirally non-invariant form of the quark current 
which we associate with $\rho^+$. Using this form of the current,
additional contributions must arise which are associated
with vector--axial-vector current mixing\footnote{We thank A. Vainshtein
for discussions on this point.}. In other words, the purely 
vectorial current is not diagonal due to the chiral anomaly, and one needs to
consider all the $\langle j_Vj_V\rangle$, $\langle j_Vj_A\rangle$,
and $\langle j_Aj_A\rangle$ correlators to obtain a well-defined
projection onto $\rh$. However, a more elegant approach would appear
to involve a direct diagonalisation of the current. To see how this
might be achieved, let us write down two forms of the 
QCD Lagrangian with an external vector source coupled to 
isovector quark current:
\begin{eqnarray}
{\cal L}_1=\cdots-m_u\bar u u -m_d \bar d d + 
\theta\fr{\alpha_s}{8\pi} G^a_{\mu\nu}\GD^a_{\mu\nu}+ V_\mu \bar u \gm d 
+  V^*_\mu \bar d \gm u
\label{l1}
\\
{\cal L}_2=\cdots -m_u\bar u u -m_d \bar d d - \theta m_*(\bar u i\gamma_5 u 
+ \bar d i\gamma_5 d ) +  V_\mu \bar u \gm d 
+  V^*_\mu \bar d \gm u
\label{l2}
\end{eqnarray} 
where the ellipses stand for the standard kinetic terms for the gauge and 
quark fields. In the absence of the external current ${\cal L}_1$ and
$  {\cal L}_2 $ are equivalent (we consider $\theta$ to be small 
and work only to linear order). The presence of the external current 
in the form written in the Eqs. (\ref{l1}) and (\ref{l2}) makes 
${\cal L}_1$ and ${\cal L}_2 $ explicitly inequivalent. The same chiral 
rotation transforms ${\cal L}_1$ to ${\cal L}_2 $ plus an extra term 
\be
{\cal L}_1 \longrightarrow {\cal L}_2+ i\theta \fr{m_d-m_u}{m_u+m_d}
(V_\mu \bar u \gm\gf d 
-  V^*_\mu \bar d \gm\gf u).
\ee
In the limit of $m_u=0$, this extra term contains $\theta$ 
explicitly which will then enter in the physical amplitudes bilinear in 
$V$. Thus we need to bear in mind that in the presence 
of $\theta$ the choice of the current 
for further use in QCD sum rules is not ``automatic'', if one wishes to
avoid mixing with other contributions. In the case of
$\rho^+$ with $m_u\neq m_d$, the $\bar d \gm u$--current should be 
used only in the basis where $\theta$ is completely rotated to 
the quark mass term. In the 
basis where $\theta$ enters in front of $G\GD$, the 
current includes additional axial-vector pieces which restore 
the correct quark mass dependence in the final answer. 
As an additional check, one 
can calculate the next order $\sim \theta^2$--corrections to CP-even 
observables and observe the dependence of the result on the choice of the 
current. 

Of course, the calculation of the electric dipole moment of 
$\rho$ does not have direct experimental implication. 
Our main motivation for calculating this quantity was to test the 
possibility of applying the QCD sum rule approach to the problem of 
EDM($\theta$).  We would like to mention here that the idea 
to consider the EDM of the $\rho$ meson resulting from the EDM of quarks 
was used in \cite{McK}. 

Returning to the problem of $d_n(\theta)$, it seems clear that there
are a number of additional difficulties which may be encountered. 
One of them refers to the correct choice of
the nucleon current at $\theta\neq 0$, as this choice can be ambiguous even 
in the normal CP-conserving case \cite{neutcurr}. Another difficulty is 
related to the necessity for a simultaneous treatment of the mass operator 
and the electromagnetic form factors. This is because in the presence of 
$\theta$ the mass operator develops an imaginary part which 
can influence the answer for $d_n(\theta)$. 
Nevertheless, this calculation appears feasible and work in this direction
is currently in progress. 
Only then can one have a reliable means to interpret 
$d_n$ directly in terms of the high-energy parameters (CP-violating phases
in the soft-breaking sector and masses of superpartners in
the case of the MSSM). 

In conclusion, 
we have demonstrated that QCD sum rules can be used for the calculation of
CP-odd electromagnetic form factors induced by the theta term. 
The result for the EDM of the vector meson, calculated in this way, 
is stable and numerically dominated by the vacuum magnetic susceptibility. 
The set of correlators which appear in the OPE part of the QCD sum rule 
were calculated via the use of the anomaly equation, in a similar manner 
to the calculation of topological susceptibility. In this way, a direct 
relation to the U(1) problem becomes apparent as the total result vanishes
in the limit $m_\eta\rightarrow m_{\pi}$. 

{\bf Acknowledgments}
We would like to thank M. Shifman and A. Vainshtein for many
valuable discussions and comments. This work was supported in part by
the Department of Energy under Grant No. DE-FG02-94ER40823.

\bibliographystyle{prsty}

\end{document}